\documentclass[prb,twocolumn,showpacs,preprintnumbers,amsmath,amssymb]{revtex4}
\usepackage{graphicx}
\usepackage{bm}

\begin{document}


\title{Unusual quantum magnetic-resistive oscillations in a superconducting structure of two circular asymmetric loops in series}


\author{V.~I.~Kuznetsov}
\email{kvi@ipmt-hpm.ac.ru}
\author{A.~A.~Firsov}
\affiliation{Institute of Microelectronics Technology and High
Purity Materials, Russian Academy of Sciences, Chernogolovka,
Moscow Region 142432, Russia}

\date{\today}

\begin{abstract}
We measured both quantum oscillations of a rectified time-averaged
direct voltage $V_{rec}(B)$ and a dc voltage $V_{dc}(B)$ as a
function of normal magnetic field $B$, in a  thin-film aluminum
structure of two asymmetric circular loops in series at
temperatures below the superconducting critical temperature
$T_{c}$. The $V_{rec}(B)$ and $V_{dc}(B)$ voltages were observed
in the structure biased only with an alternating current (without
a dc component) and only with a direct current (without an ac
component), respectively. The aim of the measurements was to find
whether interaction (nonlinear coupling) exists between quantum
magnetic-resistive states of two loops at a large distance from
each other. The distance between the loop centers was by an order
of magnitude longer than the Ginzburg-Landau coherence length
$\xi(T)$. At such distance, one would not expect to detect any
interaction between the quantum states of the loops. But we did
find such an interaction. Moreover, we found that $V_{dc}(B)$
functions (like $V_{rec}(B)$ ones) can be used to describe the
quantum states of the loops.
\end{abstract}

\maketitle
\section{INTRODUCTION}
Asymmetric superconducting loops without tunnel contacts are
interesting for a prospective technological application
\cite{sit1} and fundamental studies of unusual quantum
magnetic-field-dependent oscillations of both rectified direct
voltage \cite{sit1} and critical superconducting currents
\cite{sit2, sit3}. A superconducting asymmetric circular loop and
the asymmetric loops in series are very efficient rectifiers of ac
voltage \cite{sit1}.

Rectification effect can be interpreted as follows. If a
superconducting loop is threaded with a magnetic flux $\Phi$ and
biased by an alternating sinusoidal current (with a zero dc
component) and if, in addition, the sum of bias ac and
magnetic-filed-periodically-dependent loop circulating current
exceeds the critical current value in a certain loop part, then an
alternating voltage, with the period equal to the current period,
appears across the loop. In a strictly symmetric circular loop,
the time-averaged value of alternating voltage is zero because of
the positive voltage corresponding to certain half-periods is
cancelled out by the negative voltage corresponding to other
half-periods. In a superconducting asymmetric circular loop
\cite{sit1}, the difference between circulating current densities
in two semi-loops disturbs the symmetry between positive and
negative voltages, and a nonzero time-averaged (rectified) direct
voltage $V_{rec}(B)$ appears as a function of magnetic field $B$.

$V_{rec}(B)$ voltage oscillates with the period $\Delta
B=\Phi_{0}/S$, here $\Phi_{0}$ is the superconducting magnetic
flux quantum and $S$ is the effective loop area \cite{sit1}. In a
single asymmetric loop, $V_{rec}(B)$  is the sign-alternating
function of $B$. $V_{rec}(B)$ voltage changes its sign in fields
corresponding to integer and half-integer values of a normalized
magnetic flux $\Phi / \Phi_{0}$ \cite{sit1}.

$V_{rec}(B)$ oscillations \cite{sit1} are quite unusual. They
radically differ from oscillations in the Little-Parks (LP) effect
\cite{sit4}. As compared to the LP oscillations \cite{sit4}, in
low fields, the amplitude of the $V_{rec}(B)$ oscillations can
reach a giant magnitude that can be calculated by the expression
$I_{c}R_{N}/2 \pi$ (for a bias sinusoidal current) \cite{sit1}.
Here, $I_{c}$ is the critical superconducting current in the zero
field and $R_{N}$ is the structure resistance in the normal state.
The maximum amplitude of voltage oscillations $V_{rec}(B)$
corresponds to the maximum amplitude $\Delta R$ (from peak to
peak) of resistance oscillations that can reach $R_{N}$. For
aluminum loops \cite{sit1}, the $\Delta R$ amplitude derived from
$V_{rec}(B)$ oscillations can exceed the amplitude of
magnetoresistance oscillations expected on the basis of the LP
effect more than by one order of magnitude.

$V_{rec}(B)$ oscillations reach their extreme values (maxima and
minima) at $\Phi / \Phi_{0}$ close to $\pm (n+1/4)$, where $n$ is
an integer, while $R(B)$ oscillations in the LP effect reach their
extreme values at $\Phi / \Phi_{0}=n+1/2$.

In a single asymmetric circular loop and identical asymmetric
loops in series, a nonzero rectified voltage $V_{rec}(B)$ appears
because of a difference between critical superconducting currents
$I_{c+}(B)$ and $I_{c-}(B)$ measured \cite{sit2, sit3} for
arbitrarily positive and negative half-waves of a bias ac.

$I_{c+}(B)$ and $I_{c-}(B)$ oscillations were unexpectedly found
to be unusual. It was revealed that in low fields the curves
$I_{c+}(B)$ and $I_{c-}(B)$ are shifted from one another with a
$\pi$ phase shift (corresponds to magnetic-field shift of
$\Phi_{0}/2$) \cite{sit2, sit3}. Note that according to
contemporary theoretical conceptions this incomprehensible $\pi$
phase shift can be hardly expected in the studied asymmetric
structures \cite{sit1, sit2, sit3}.

Moreover, like $V_{rec}(B)$ oscillations these $I_{c+}(B)$ and
$I_{c-}(B)$ oscillations reach their extreme values at $\Phi /
\Phi_{0}$ close to $\pm (n+1/4)$. Some other striking features of
$V_{rec}(B)$, $I_{c+}(B)$, and $I_{c-}(B)$ oscillations were also
found in asymmetric loops \cite{sit2, sit3}.

Unlike LP oscillations, the unusual quantum $V_{rec}(B)$,
$I_{c+}(B)$, and $I_{c-}(B)$ oscillations \cite{sit1, sit3} in an
asymmetric circular loop cannot be explained in the framework of
the simple Ginzburg-Landau (GL) quasi-one-dimensional model
\cite{sit5}, using only the requirement of superconducting fluxoid
quantization \cite{sit6}. Thus, experimental studies of the
unusual quantum oscillations \cite{sit1, sit2, sit3, sit7, sit8}
in simple superconducting circular-asymmetric structures generate
many unsolved questions.

Besides the rectification of ac voltage, an asymmetric loop is
interesting for other applications. Earlier \cite{sit1, sit7,
sit8}, it was assumed that the quantum states of a single
superconducting asymmetric circular loop placed in the normal
magnetic field $B$ at $T$ below $T_{c}$ can be described by the
oscillating superconducting circulating current of the loop
$I_{R}(B)$. In order to determine the time-averaged quantum
states, the loop was periodically switched from superconducting
state to resistive state and back by a bias alternating current
(without a dc component) with an amplitude close to the critical
current value. As a result of the multiple switching, the
rectified direct voltage $V_{rec}(B)$ appears in the loop
\cite{sit1}. At certain values of the magnetic field, $V_{rec}(B)$
can be directly proportional to $I_{R}(B)$ in a single asymmetric
circular loop \cite{sit1}. Therefore, measurements of $V_{rec}(B)$
can allow the readout of the time-averaged quantum states of the
loop \cite{sit1}. Moreover, the oscillating $V_{rec}(B)$ voltage
recorded at the different current values can describe quantum
magneto-resistive states of the loop (the states depend both on
the magnetic field and the bias ac). Quantum magneto-resistive
states of the two directly connected asymmetric circular loops
\cite{sit7, sit8} can be described by quantum magneto-resistive
states of each loop and coupling between the states of the loops.

In addition, an asymmetric loop of high-resistance material with
an extremely small wall narrowing could be used as an element of a
superconducting flux qubit \cite{sit9, sit10} with quantum
phase-slip centers \cite{sit11, sit12}. Two successive loops of
this kind could be an analog of two successive flux qubits.

For certain technological applications, it is necessary to know
both the strength and mechanism of coupling between quantum states
of loops. Earlier, an interaction (nonlinear coupling) was
revealed \ between quantum magnetic-resistive states of two
different superconducting directly connected asymmetric circular
loops forming a figure-of-eight-shaped structure \cite{sit7,
sit8}. To determine the quantum state of each loop and coupling
between the loops, rectified voltage $V_{rec}(B)$ was measured in
a figure-of-eight-shaped structure pierced with a magnetic flux
and biased with a low-frequency current (without a dc component)
and with an amplitude close to critical, at $T$ slightly below
$T_{c}$ \cite{sit7, sit8}. Possible mechanisms of the interaction
between the loops can be magnetic coupling and electro-dynamic
coupling through a bias ac.

We assume that electro-dynamic coupling between two successive
loops that occurs through a bias ac is a nonlocal phenomenon with
the nonlocal superconducting length \cite{sit13, sit14, sit15}
close, by the order of magnitude, to the Ginzburg-Landau coherence
length $\xi(T)$ \cite{sit6}. Note that the nonlocal length
estimated from Ref. \cite{sit16} can reach the value several times
exceeding $\xi(T)$.

Nonlinear \ coupling \ between quantum magnetic-resistive states
of two asymmetric loops in series should become weaker with an
increasing distance between them. It can be assumed that the most
long-scaled electro-dynamic coupling should almost disappear if
the spacing between loop centers increases to $10 \xi(T)$.

\begin{figure}
\includegraphics[width=1.0\columnwidth]{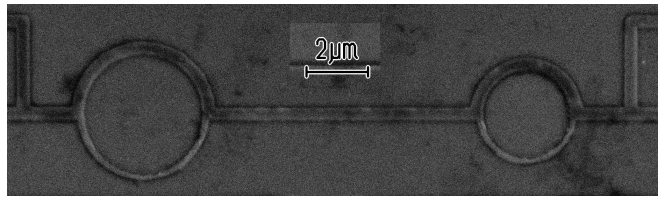}
\caption{\label{image} SEM image of the structure. The scale bar:
$2$ $\mu$m.}
\end{figure}

The aim of this work was to find the largest distance between the
loops at which the coupling between quantum magnetic-resistive
states of the loops would still occur. For this purpose, we
experimentally studied the quantum magnetic-resistive behavior of
two different superconducting aluminum asymmetric circular loops
connected in series with a wire of a length (Fig. \ref{image})
close to the penetration depth of a nonuniform electric field into
a superconductor \cite{sit5, sit17} $\Lambda_{E}$.

Like the authors of Refs. \cite{sit7, sit8}, we measured the
rectified direct voltage $V_{rec}(B)$ in the structure (Fig.
\ref{image}) versus normal magnetic field $B$ and a bias
sinusoidal low-frequency current (without a dc component) at
temperatures $T$ slightly below $T_{c}$ in order to determine the
quantum magnetic-resistive states of two loops in series and the
coupling between the loop states. In addition, we measured a dc
voltage $V_{dc}(B)$ as a function of $B$ and bias dc (without an
ac component) through the structure.

One more goal was to test whether $V_{dc}(B)$ oscillations can be
used to describe the quantum magnetic-resistive states of an
asymmetric structure. We also made a comparison of $V_{rec}(B)$
and $V_{dc}(B)$ oscillations.

\section{SAMPLES AND EXPERIMENTAL PROCEDURE}
A 45 nm thick structure of two loops connected in series was
fabricated by thermal sputtering of aluminum onto a silicon
substrate using the lift-off process of electron-beam lithography.
Figure \ref{image} displays a scanning electron microscopy (SEM)
image of the structure. It consists of two different successive
asymmetric circular loops with the distance between the loop
centers $L=12.5$ $\mu$m. The average widths of all narrow and wide
wires in the sample central part are $w_{n}=0.22$ $\mu$m  and
$w_{w}=0.41$ $\mu$m, respectively. The circular asymmetry permits
the observation of nonzero rectified voltage $V_{rec}(B)$ in the
structure \cite{sit1, sit7, sit8}. The minimum area of the loop is
the internal area within the inner loop border. The minimum areas
of the larger and smaller loops are $S_{Lmin}=11.57$ $\mu\rm
m^{2}$ and $S_{Smin}=6.34$ $\mu\rm m^{2}$, respectively. From the
structure geometry, the average areas of the larger and smaller
loops are $S_{Lg}=13.93$ $\mu\rm m^{2}$ and $S_{Sg}=7.92$ $\mu\rm
m^{2}$, respectively.

The structure has the following parameters. The critical
superconducting temperature $T_{c}=1.355\pm 0.001$ K was
determined from the midpoint of normal-superconducting transition
$R(T)$ in a zero field. The total normal-state resistance measured
between two vertical wires at $T=4.2$ K is $R_{N}=32$ $\Omega$.
The ratio of room-temperature to the helium-temperature resistance
is $R_{300}/R_{4.2}=2$. Sheet resistance is $R_{S}=0.69$ $\Omega$,
the resistivity is then $\rho=3.105\times 10^{-8}$ $\Omega$ m.
From the expression \cite{sit1, sit18} $\rho l=(6 \pm 2)\times
10^{-16}$ $\Omega$ $\rm m^2$, we determine the electron mean free
path $l=19$ nm. The superconducting coherence length of pure
aluminum at $T=0$ is $\xi_{0}=1.6$ $\mu$m. Hence, the structure is
a "dirty" superconductor, because $l \ll \xi_{0}$. Therefore, for
this structure the temperature-dependent superconducting G-L
coherence length at temperatures slightly below $T_{c}$ is
determined from the expression \cite{sit6, sit19}
$\xi(T)=\xi(0)(1-T/T_{c})^{-1/2}$, where
$\xi(0)=0.85(\xi_{0}l)^{1/2}=0.15$ $\mu$m. In the studied
temperature range, $\xi(T)=0.85-1.2$ $\mu$m. For this structure,
the theoretical estimation results in the penetration depth of a
nonuniform electric field into a superconductor \cite{sit5, sit17}
$\Lambda_{E} \approx 10$ $\mu$m in the experimental temperature
range. So, $\Lambda_{E}$ is larger than $\xi(T)$ by an order of
magnitude.

Two types of four-probe measurements of voltage oscillations
versus magnetic field normal to the substrate surface were
performed in the structure (Fig. \ref{image}). In the first case,
a rectified time-averaged direct voltage $V_{rec}(B)$ appeared in
the structure biased by a sinusoidal current (without a dc
component) $I_{ac}(t)=I_{ac}\sin(2\pi\nu t)$, with the amplitude
$I_{ac}$ close to the critical current in the zero field $ I_{c}$
at frequencies $\nu$ from $1$ to $10$ kHz at $T$ slightly below
$T_{c}$. The experimental procedure was similar to that in Refs.
\cite{sit1, sit7, sit8}. The $V_{rec}(B)$ voltage was measured at
a slowly varying magnetic field. The $V_{rec}(B)$ voltage was the
time-averaged value of the alternating voltage $V_{ac}(B, t)$ over
a time interval $\Delta t$, i.e. $V_{rec}(B)=\frac{1}{\Delta
t}\int^{\Delta t}_{0} V_{ac}(B, t)dt$. The condition $\Delta t
>20\Delta t_{I}$ was valid ($\Delta t_{I}$ is the period of bias
ac).

In the second case, a dc voltage $V_{dc}(B)$ was measured when a
bias direct current $I_{dc}$ (without an ac component) passed
through the structure. As well as $V_{rec}(B)$ curves, $V_{dc}(B)$
data were obtained at different dc values, close to the $I_{c}(T,
B=0)$ at $T$ slightly below $T_{c}$.

In addition, we measured a dc voltage $V_{dc}(I)$ as a function of
the bias dc at different magnetic-field values.

\section{RESULTS AND DISCUSSION}
\subsection{Main results}
\begin{figure*}\includegraphics[width=3.35 in]{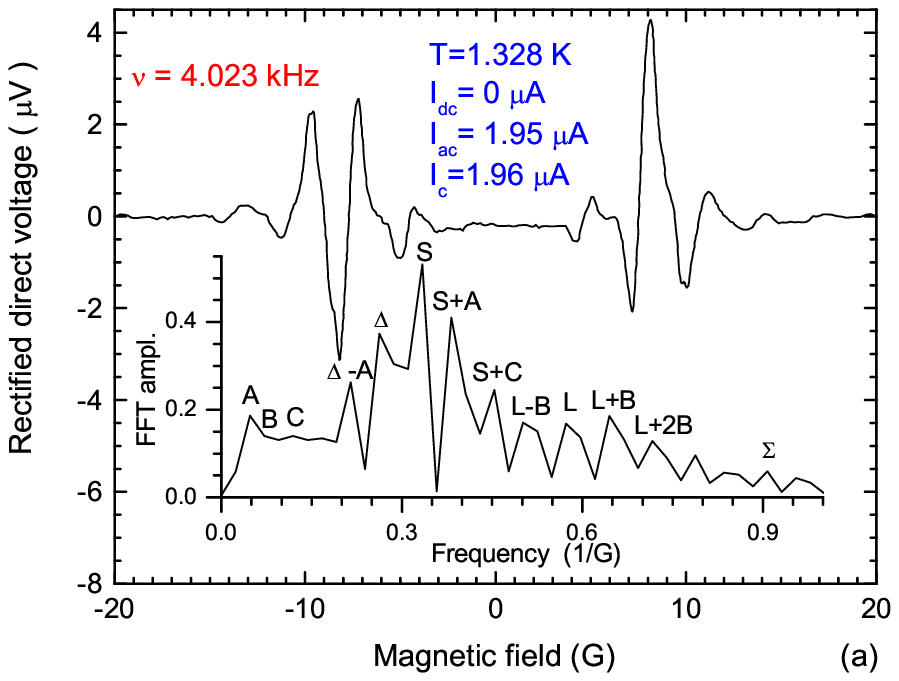}\includegraphics[width=3.35 in]{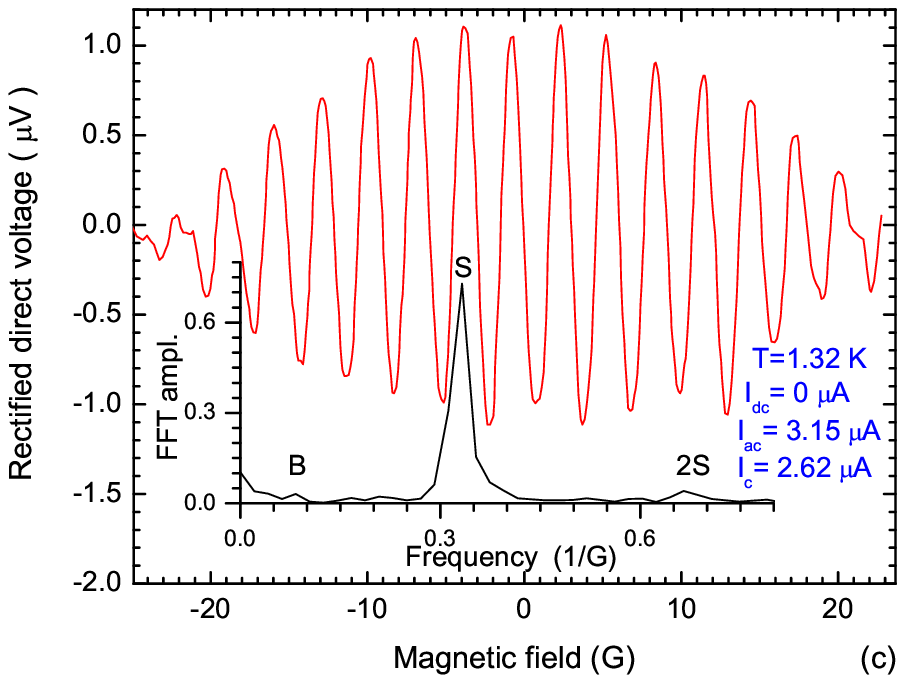}
\includegraphics[width=3.35 in]{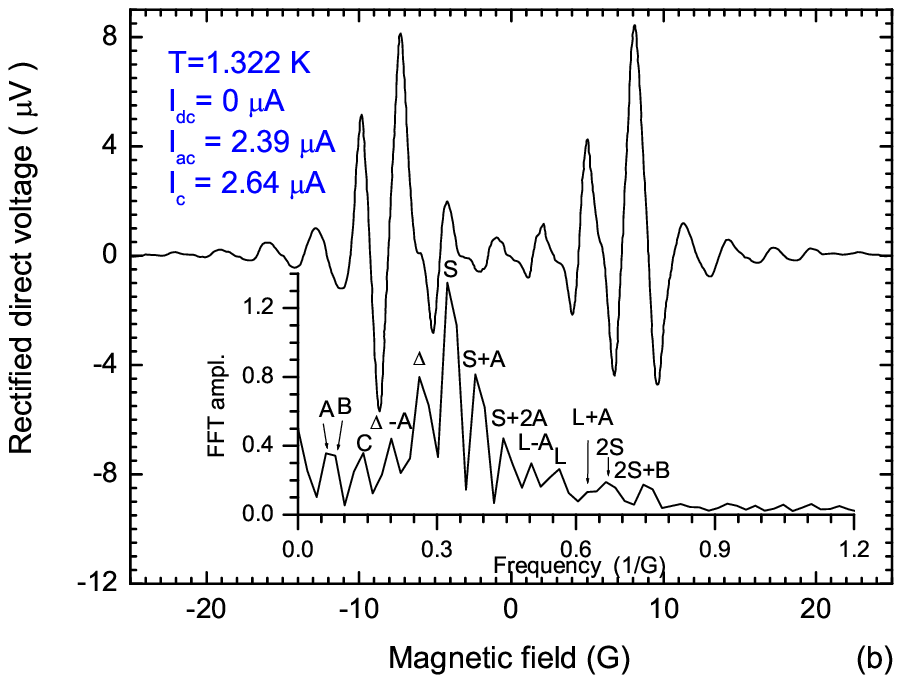}\includegraphics[width=3.35 in]{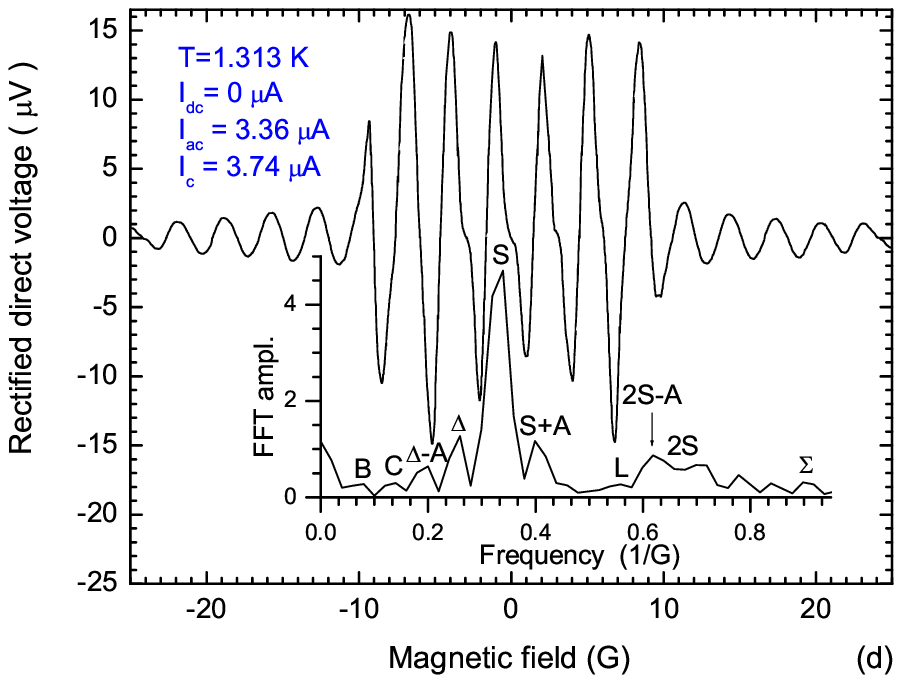}
\caption{\label{voltage} (Color online) [(a)-(d)] $V_{rec}(B)$
curves measured at the parameters shown in the figure. Magnetic
field values are given in G $=10^{-4}$ T. Insets: FFT spectra of
the curves. Symbols $S$, $L$, $\Delta$, $\Sigma$, and $2S$ show
the spectral regions corresponding to the fundamental frequencies
of the smaller and larger loops, difference and sum of these
frequencies, and second higher harmonics of the fundamental
frequency of the smaller loop, respectively. Symbols $A$, $C$,
$B$, and $D$ mark the regions corresponding to the extra low
frequencies $f_{A}=0.048-0.060$ ${\rm G}^{-1}$,
$f_{C}=0.119-0.141$ ${\rm G}^{-1}$, their difference
$f_{B}=f_{C}-f_{A}=0.072-0.081$ ${\rm G}^{-1}$, and their sum
$f_{D}=f_{C}+f_{A}=0.167-0.201$ ${\rm G}^{-1}$, respectively.
Linear combinations of symbols $S$, $L$, $\Delta$, $\Sigma$, $A$,
$C$, $B$, and $D$ denote various combination frequencies. For
example, the sum $S+A$ near a peak shows that the peak corresponds
to the summation frequency $f_{S}+f_{A}$.}
\end{figure*}

\begin{figure}\includegraphics[width=3.35 in]{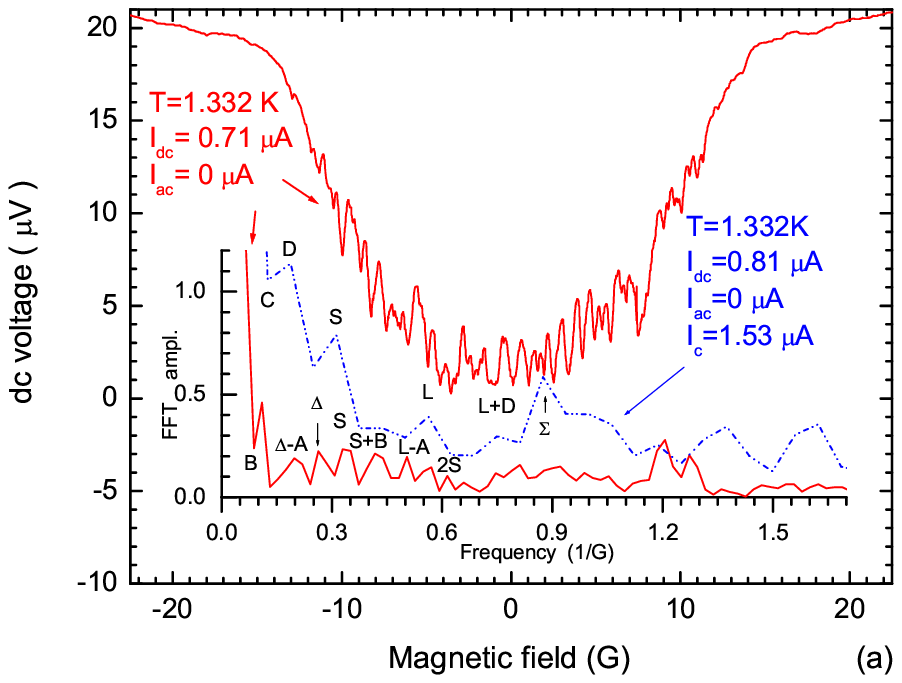}
\includegraphics[width=3.35 in]{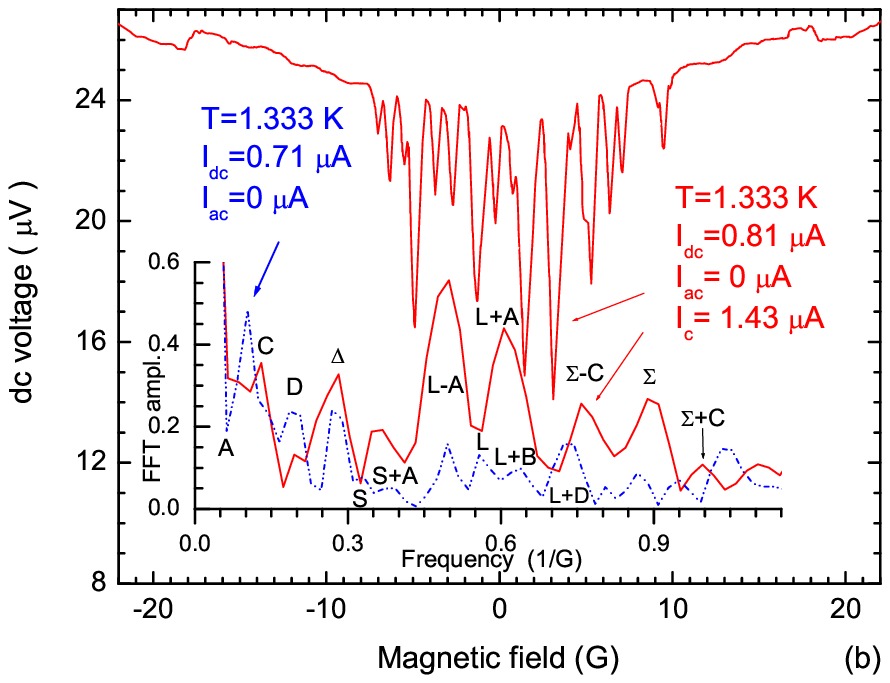}
\includegraphics[width=3.35 in]{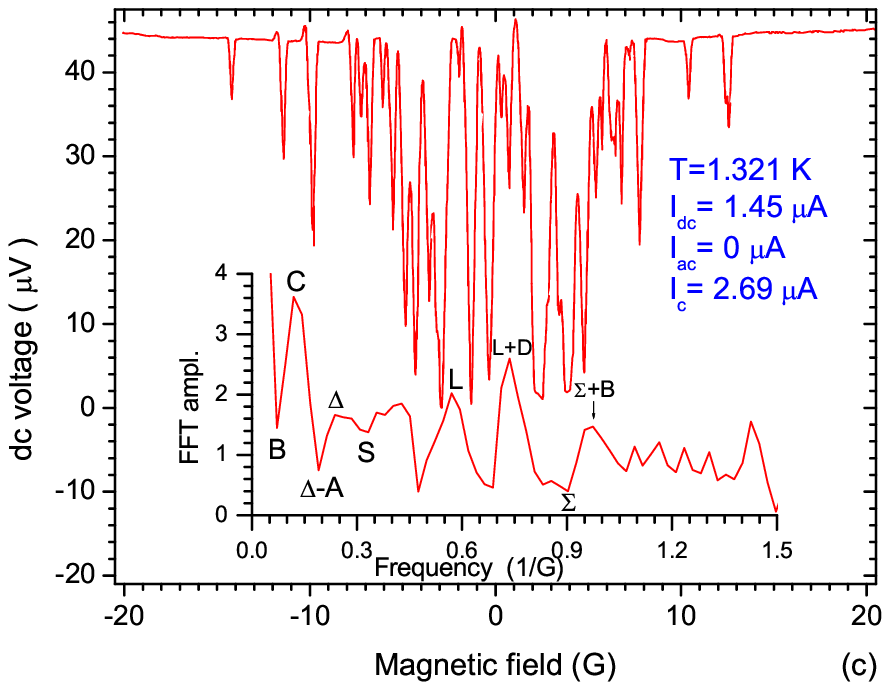}
\caption{\label{volt} (Color online) [(a)-(c)] $V_{dc}(B)$ curves
measured at the parameters shown in the figure. Field values are
given in G $=10^{-4}$ T. Insets: solid lines correspond to FFT
spectra of the curves, dash-dot-dot lines present spectra of other
$V_{dc}(B)$ curves (not shown in the figure) measured at the
parameters shown in the insets of Figs. \ref{volt}(a) and
\ref{volt}(b). Regions corresponding to certain frequencies are
marked with symbols and their combinations the way used in Fig.
\ref{voltage}. From the insets, extra low frequencies are equal:
$f_{A}=0.050-0.062$ ${\rm G}^{-1}$, $f_{C}=0.120-0.130$ ${\rm
G}^{-1}$, $f_{B}=0.070-0.088$ ${\rm G}^{-1}$, and
$f_{D}=0.166-0.187$ ${\rm G}^{-1}$.}
\end{figure}

The measured $V_{rec}(B)$ and $V_{dc}(B)$ oscillations are shown
in Figs. \ref{voltage} and \ref{volt}, respectively. The values of
the direct $I_{dc}$, alternating $I_{ac}$ and critical currents
$I_{c}$ together with the temperature $T$ are given in the
figures. For a detailed study, fast Fourier transformations (FFT)
of the $V_{rec}(B)$ and $V_{dc}(B)$ functions were obtained using
4096 uniformly of distributed points in magnetic fields from -25
to +25 G (Gauss $=10^{-4}$ Tesla). The Fourier spectra of these
functions exhibit a variety of peaks of different magnitudes and
frequencies (insets of Figs. \ref{voltage} and \ref{volt}). The
frequency $f$ has the meaning of a value inversely proportional to
a certain period of voltage oscillations, i.e. $f=1/ \Delta B$.
The great number of various frequencies in the spectra indicates
the presence of various periodic magnetic-field-dependent
responses of the structure.

Fundamental frequencies of the smaller and larger loops are
\begin{equation}
f_{S}=1/\Delta B_{S}=S_{S}/\Phi_{0}~,~~\rm ~~f_{L}=1/\Delta
B_{L}=S_{L}/\Phi_{0}~,
\end{equation} respectively.
Here $\Delta B_{S}$ and $\Delta B_{L}$ are the periods of voltage
oscillations in the loops,  and $S_{S}$ and $S_{L}$ are effective
loop areas, with $S$ and $L$ relating to the smaller and larger
loops, respectively. The expected fundamental frequencies
corresponding to averaged geometrical areas of the loops are
$f_{Sg}=0.38$ ${\rm G}^{-1}$ and $ f_{Lg}=0.67$ ${\rm G}^{-1}$.
The minimum expected values of fundamental frequencies of the
loops are $f_{Smin}=0.31$ ${\rm G}^{-1}$ and $ f_{Lmin}=0.56$
${\rm G}^{-1}$, corresponding to the minimum areas of the smaller
and larger loops.

Figure \ref{voltage} presents the rectified direct voltage
$V_{rec}(B)$ versus magnetic field through the structure. The
structure was biased by a sinusoidal current with the frequency
$\nu=4.023$ kHz (without a dc component) and a current amplitude
$I_{ac}$ close to critical $I_{c}$ at $T$ slightly below $T_{c}$.

Figure \ref{volt} shows a dc voltage $V_{dc}(B)$ through the
structure biased with direct current (without any ac component) at
$T$ slightly below $T_{c}$. The FFT spectra of both $V_{rec}(B)$
and $V_{dc}(B)$ functions exhibit peaks (labeled $S$ and $L$)
localized near the fundamental frequencies $f_{S}=0.31-0.33$ ${\rm
G}^{-1}$ and $ f_{L}=0.56-0.57$ ${\rm G}^{-1}$, corresponding to
the smaller and larger loops. These obtained $f_{S}$ and $f_{L}$
frequencies are close to the corresponding minimum expected
fundamental $f_{Smin}$ and $f_{Lmin}$ frequencies of the loops
[insets of Figs. \ref{voltage}(a), \ref{voltage}(b),
\ref{voltage}(d), \ref{volt}(a), \ref{volt}(b)]. This means that
the effective loop areas are close to the corresponding minimum
areas of the loops.

The spectral peaks labeled $2S$ at $f_{S2}=2f_{S}$ correspond to
the second higher harmonics of $f_{S}$ frequency [insets of Figs.
\ref{voltage}(b), \ref{voltage}(c), and \ref{volt}(a)].

Apart from fundamental $f_{S}$ and $f_{L}$ frequencies, the
spectra exhibit the difference frequency $f_{\Delta}= f_{L}-f_{S}$
[insets of Figs. \ref{voltage}(a), \ref{voltage}(b),
\ref{voltage}(d) and \ref{volt}(a) - lower curve, \ref{volt}(b)
and \ref{volt}(c)] and the summation frequency
$f_{\Sigma}=f_{L}+f_{S}$ [insets of Figs. \ref{voltage}(a),
\ref{voltage}(d), \ref{volt}(a)- both curves and \ref{volt}(b)].
In the insets of Figs. \ref{voltage} and \ref{volt}, the symbols
$\Delta$ and $\Sigma$ denote peaks corresponding to $f_{\Delta}$
and $f_{\Sigma}$ frequencies.

These difference and summation frequencies $f_{\Delta}=
f_{L}-f_{S}$ and $f_{\Sigma}=f_{L}+f_{S}$ point out to the
interaction (nonlinear coupling) between the loops. Earlier,
$f_{\Delta}$ and $ f_{\Sigma}$ frequencies were found in the
Fourier spectrum of $V_{rec}(B)$ voltage in a
figure-of-eight-shaped structure \cite{sit7, sit8}.

The spectra also demonstrate extra peaks of low frequencies
$f_{A}=0.048-0.062$ ${\rm G}^{-1}$, $f_{C}=0.119-0.141$ ${\rm
G}^{-1}$ and difference $f_{B}= f_{C}-f_{A}=0.070-0.088$ ${\rm
G}^{-1}$, and summation $f_{D}= f_{C}+f_{A}=0.166-0.201$ ${\rm
G}^{-1}$ frequencies (insets of Figs. \ref{voltage} and
\ref{volt}) denoted as $A$, $C$, $B$, and $D$, respectively. We
found that $1/f_{C}$ and $1/f_{A}$ are several times smaller than
the critical magnetic field at which the superconducting order
parameter is radically suppressed in the structure.

Moreover, combination satellite frequencies $f_{S}+f_{A}$,
$f_{S}+2f_{A}$, $f_{S}+f_{B}$, $f_{S}+f_{C}$ localized to the
right of the $f_{S}$ frequency, $f_{L} \pm f_{A}$, $f_{L} \pm
f_{B}$, $f_{L}+2f_{B}$, $f_{L}+f_{D}$ near $f_{L}$ frequency,
$f_{\Delta}-f_{A}$ to the left of $f_{\Delta}$ and $f_{\Sigma} \pm
f_{C}$, $f_{\Sigma}+f_{B}$ near $f_{\Sigma}$ can be seen in the
spectra [insets of Figs. \ref{voltage}(a), \ref{voltage}(b),
\ref{voltage}(d), \ref{volt}(a), \ref{volt}(b) and \ref{volt}(c)].

Earlier, extra low frequencies and satellite frequencies which are
combinations of the loop fundamental frequency and a low frequency
were found in the Fourier spectrum of the of $V_{rec}(B)$
oscillations in a single almost symmetric circular loop
\cite{sit20}. At present, the reason of the low frequencies
appearance in the spectra is not found.

Relative and absolute resistive contributions of both loops into
$V_{rec}(B)$ and $V_{dc}(B)$ oscillations vary with external
parameters. The resistive response of the smaller loop $S$ is
often larger than that of the larger loop $L$ [insets of Figs.
\ref{voltage}(a)-(d) and \ref{volt}(a)]. At $T$ slightly below
$T_{c}$, a larger response of the larger loop $L$ was observed as
compared with the smaller loop response $S$ [insets of Figs.
\ref{volt}(b) - the lower curve and \ref{volt}(c)].

With changing external parameters, dips were observed instead of
peaks at certain frequencies. For example, dips are observed near
the $f_{S}$ and $f_{L}$ frequencies [upper curve in the inset of
Fig. \ref{volt}(b)] instead of peaks $S$ and $L$ [lower curve in
the inset of Fig. \ref{volt}(b)]. A dip $S$ at $f_{S}$ is also
seen in the inset of Fig. \ref{volt}(c). The dip in the inset of
Fig. \ref{voltage}(d) labeled $2S$ corresponds to the second
higher harmonics of the $f_{S}$ frequency. The inset of Fig.
\ref{volt}(c) exhibits a dip $\Sigma$ near the summation frequency
$f_{\Sigma}$. Instead of the spectral peak $\Delta$ observed at
the frequency close to $f_{\Delta}$ at current $I_{dc}=0.71$
$\mu$A [inset of Fig. \ref{volt}(a), lower curve], a dip $\Delta$
appears at $I_{dc}=0.81$ $\mu$A [inset of Fig. \ref{volt}(a),
upper curve]. Both curves [Fig. \ref{volt}(a)] were measured at
the same temperature. With increasing current a dip $\Delta-A$
[inset of Fig. \ref{volt}(c)] appears instead of the extra
low-frequency spectral peak $\Delta-A$ [inset of Fig.
\ref{volt}(a)].

We believe that dips arise because of the nonlinear coupling of
loop (wire) oscillations to give the product of two or more
periodic signals with certain frequencies (amplitude modulation).
In an ordinary case, an amplitude modulation is a periodic change
in the amplitude of high frequency carrier oscillations by a low
modulating frequency. Instead of the original frequencies, the
spectrum of this product would contain the difference and
summation frequencies. For example, the upper spectrum in the
inset of Fig. \ref{volt}(b) exhibits a dip at the frequency
$f_{L}$ and two side satellite peaks at frequencies $f_{L-A}$ and
$f_{L+A}$ instead of the peak at the $f_{L}$. At the same time,
apart from the side satellites at $f_{L-B}$ and $f_{L+B}$, Fig.
\ref{voltage}(a) shows a peak at the frequency $f_{L}$ suggesting
the presence of the carrier signal in the spectrum.

So, we consider that both the $V_{rec}(B)$ and $V_{dc}(B)$
oscillations can allow us to determine the quantum
magneto-resistive states of the 2-loop structure and the coupling
between the quantum states.

In order to obtain a supplementary information, we measured
$V_{dc}(I)$ curves (not presented here) in different magnetic
fields for two opposite directions of the bias dc sweep. In low
fields, a large hysteresis of the $V_{dc}(I)$ curves is observed
even at temperatures sufficiently close to $T_{c}$ and low
currents $I_{dc}<1$ $\mu$A. We believe that the hysteresis is due
to a quasiparticle overheating \cite{ sit16} that can be caused by
the energy dissipation in thermally activated phase-slip centers
(TAPSCs) \cite{sit5, sit16, sit17}. TAPSC generates a
quasiparticle imbalance and consequently, a nonuniform electric
field in a nonequilibrium region with the size close to
$2\Lambda_{E}$.

\subsection{Nonlinear coupling, combination frequencies, heating effects}

The difference and sum of fundamental frequencies in FFT spectra
suggest that nonlinear coupling does exist between quantum
magnetic-resistive states of the successive loops. Let consider
now how nonlinear coupling between loops arises and how
combination frequencies appear. Nonlinear coupling between
directly connected loops can be due to magnetic inductive coupling
between loops and electrodynamic interaction through a bias
current \cite{sit7, sit8}. However, magnetic coupling cannot
explain the great number of frequencies observed in the
oscillation spectra, e.g. the sum of loop fundamental frequencies.
We assume that in the case of successive loops, the inductive
coupling between loops would become much weaker as the distance
between the loops increases. Then, the interaction through a bias
current should predominate over the inductive coupling in the
studied 2-loop structure.

Quantum magnetic-resistive properties of a single loop depend both
on the loop circulating current and the bias current. A fraction
of the bias current passing through the single loop becomes
magnetic-field-dependent, oscillating with a period equal to that
of the loop circulating current oscillations ($\Delta
B_{S}=1/f_{S}$ for the smaller loop and $\Delta B_{L}=1/f_{L}$ for
the larger loop). Because of the finite spatial change of
superconducting order parameter, the nonlocal the $V_{rec}(B)$ or
$V_{dc}(B)$ oscillations can be expected to appear on a wire part
located outside of the loop at a distance equal to the nonlocal
superconducting length. As shown experimentally and theoretically
in Refs. \cite{sit13, sit14, sit15}, the nonlocal superconducting
length is close to the $\xi (T)$ length.

If the length of a wire connecting two successive loops is shorter
than $\pi \xi (T)$, then it can be expected that some fraction of
the oscillating current leaving one of the loops could also pass
through the other loop in the 2-loop structure. As a result,
current (voltage) oscillations with the fundamental frequency
$f_{S}$ corresponding to the smaller loop could be amplitude
modulated by the oscillations with the fundamental frequency
$f_{L}$ corresponding to the larger loop and vice versa. So, the
oscillating current fraction passing through both loops becomes
dependent on both loop oscillating currents. As a result, a
nonlinear coupling occurs between quantum magneto-resistive states
of the loops. The amplitude modulation i.e. the multiplication of
one oscillating signal by another oscillating signal should be
expected to lead to the appearance of combination frequencies.

Note that if the distance between the two loops is several times
larger than $\pi \xi (T)$, then one can hardly expect to observe
the coupling between the quantum states of the loops. In the
2-loop structure, the distance between the loop centers is close
to $13 \xi (T)$, therefore the coupling can be expected to
disappear. Nevertheless, we found the coupling.

Here we provide an explanation for the unexpected coupling. We
consider that the 2-loop structure is in a nonequilibrium state
with a nonequilibrium length $\Lambda_{E}$. Then, weak
$V_{rec}(B)$ or $V_{dc}(B)$ oscillations can be expected to appear
on a wire part located outside of the loop at the distance several
times larger than $\xi (T)$. In addition, the weak coupling
between the quantum states of two loops due to a nonlocal effect
can be expected to be observed in the 2-loop structure.

In a nonequilibrium state at distances between the loop centers
close to $\Lambda_{E}$, the coupling can be still noticeable
whereas the inductive coupling between the loops should almost
disappear. Therefore, it is most likely that the coupling of
quantum magnetic-resistive states of both loops mainly occurs
through a common bias ac (dc) owing to the nonlocal effect.

Let us speculate how the quasiparticle overheating can effect on
the quantum magneto-resistive behavior of the 2-loop structure. As
a result of the quasiparticle overheating (an increase in the
effective local quasiparticle temperature) \cite{ sit16} of the
structure nonequilibrium region, an effective nonequilibrium
length $\Lambda_{E}$ should be expected to increase. Since the
effective resistance of the region is directly proportional to
$\Lambda_{E}$, then, one would expect that the amplitude of the
magneto-resistive oscillations would be increased. Moreover, we
believe that the quasiparticle overheating can result in an
increase in the coupling between the loops. If the overheating
would be every strong and the quasiparticle temperature would
exceed $T_{c}$ in the immediate vicinity of the $2\Lambda_{E}$
region, then the nonequilibrium region should be expected to
transform to the normal-state region with a total size larger then
$2\Lambda_{E}$ and with an increased resistance. Moreover, the
strong overheating periodically driven by the field would result
in the giant amplitude of the quantum oscillations that is due to
the switching between a state close to the normal state and a
state close to the superconducting state. So, we assume that the
overheating not only doesn't weaken quantum oscillations, but even
can strengthen the oscillations. Moreover, it is possible that the
overheating can even result in an increase in the loop coupling.

\subsection{$V_{rec}(B)$ oscillations}

The amplitude of $V_{rec}(B)$ oscillations is the function of both
a bias ac amplitude $I_{ac}$ and a magnetic filed. As was noted
above, in low fields $V_{rec}(B)$ oscillations in a single
asymmetric circular loop (in a system of identical loops in
series) \cite{sit1} are of unusual character. For example, they
have a giant amplitude at a bias ac amplitude $I_{ac}$  close to
the critical value. In the studied structure, the behavior of
$V_{rec}(B)$ oscillations is also unusual in a certain region of
low fields. In this region, the oscillation amplitude is maximum
and almost independent of the magnetic field [Fig.
\ref{voltage}(d)]. Outside the region, the oscillations first
drastically decrease with increasing field and then smoothly fade
in high fields [Fig. \ref{voltage}(d)].

In low fields, the weak magnetic-field dependence of the
oscillation amplitude is probably caused by the overheating of the
nonequilibrium region.

In a general case, in low magnetic fields, the $V_{rec}(B)$
oscillations cannot be described \cite{sit3} in the framework of
the simple GL quasi-one-dimensional theory \cite{sit5, sit17}, if
only the requirement of superconducting fluxoid quantization
\cite{sit6} is used.

The maximum value of $V_{rec}(B)$ oscillations in the structure as
well as and the maximum magneto-resistive response of a single
asymmetric loop are determined by that how close the resistive
state of the structure part (loop) is to the midpoint of the
superconducting-normal ($S$-$N$) transition \cite{sit7}.

On transition from a state close to the superconducting state to a
state more close to the midpoint of the $S$-$N$ transition, the
amplitude of $V_{rec}(B)$ oscillations increases in fields close
to zero [Figs. \ref{voltage}(a) and \ref{voltage}(b)]. When the
condition $I_{ac}\approx I_{c}$ holds, these oscillations reach
their maxima in fields close to zero, with the state corresponding
to the midpoint of the $S$-$N$ transition being realized [Fig.
\ref{voltage}(d)]. When $I_{ac}>I_{c}$, a state more close to the
normal realizes [Fig. \ref{voltage}(c)]. The oscillations also
reach their maxima in fields close to zero. The oscillation
amplitude, however, decreases [Fig. \ref{voltage}(c)]. The
magneto-resistive response of the smaller loop $S$ dominates over
the larger loop response $L$ at the parameters of Fig.
\ref{voltage}. The spectral peak corresponding to the larger loop
$L$ practically disappears with increasing current [inset of Fig.
\ref{voltage}(c)].

\subsection{Comparison of $V_{rec}(B)$ and $V_{dc}(B)$ oscillations}

1. Let compare the $V_{rec}(B)$ and $V_{dc}(B)$ oscillations and
their spectra in the 2-loop structure (Fig. \ref{image}). We found
both the $V_{rec}(B)$ and $V_{dc}(B)$ voltages measured in the
structure under applied ac (with a zero dc component) and dc (with
a zero ac component) respectively, can give information about the
structure quantum magneto-resistive states and nonlinear coupling
between the loops.

Although the $V_{rec}(B)$ and $V_{dc}(B)$ functions differ
fundamentally, their spectra contain a similar set of frequencies.
The spectra show quantitative and relative differences between
periodic magnetic-field responses of the structure biased with ac
(with a zero dc component) and dc (with a zero ac component) at
the same values of $T$ and bias current $I_{ac}(I_{dc})$. The
response of the smaller loop often dominated over the one of the
larger loop when $V_{rec}(B)$ was measured (Fig. \ref{voltage}).
During $V_{dc}(B)$ measurements, a response of any of the loops
could be dominating at certain values of $T$ and $I_{dc}$.
Moreover, unlike $V_{rec}(B)$ oscillations, the $V_{dc}(B)$ ones
reach their maxima in low fields at somewhat smaller currents at
the same $T$ (Figs. \ref{voltage} and \ref{volt}).

2. Now compare the $V_{rec}(B)$ and $V_{dc}(B)$ oscillations in
the structure with $V_{rec}(B)$ oscillations in the
figure-of-eight-shaped structure of Ref. \cite{sit7}. Unlike the
spectra of $V_{rec}(B)$ oscillations in the figure-of-eight-shaped
structure, the spectra of $V_{rec}(B)$ and $V_{dc}(B)$
oscillations in the structure virtually do not contain higher
harmonics of loop fundamental frequencies expect for the second
harmonics of $f_{S}$. Moreover, the spectra of $V_{rec}(B)$ and
$V_{dc}(B)$ oscillations in the structure exhibit extra peaks at
low frequencies and peaks corresponding to extra combination
frequencies. The difference is most appears due to another
geometry of the structure as compared to that in Ref. \cite{sit7}.

It is seen that $V_{rec}(B)$ and $V_{dc}(B)$ oscillations can be
dependent on the field direction. The magnetic asymmetry (parity
and oddness violations of $V_{dc}(B)$ and $V_{rec}(B)$ functions)
considerably exceeds an experimental error. The asymmetry
practically disappears at high bias current values [Fig.
\ref{voltage}(c)]. A similar magnetic asymmetry was first observed
in the figure-of-eight-shaped structure \cite{sit8}. We speculate
that a possible reason for the asymmetry is in an increase in the
termo-dynamical instability in the structure (the system of the
TAPSCs) that can be caused by the quasiparticle overheating. A
careful experimental study of the quantum oscillations in a single
asymmetric loop would provide us with clues for an acceptable
explanation of the asymmetry.

3. Compare the $V_{rec}(B)$ and $V_{dc}(B)$ oscillations and their
spectra in the structure with LP oscillations in a hollow
thin-walled superconducting cylinder. In a low field region, the
oscillations in the structure and those in a single asymmetric
loop \cite{sit1} are of unusual character. The striking difference
of $V_{dc}(B)$ oscillations in the structure from LP oscillations
is clearly seen in Fig. \ref{volt}(c). In a low field region, the
peak-to-peak amplitude of $V_{dc}(B)$ oscillations can reach a
value close to the total voltage through the whole structure in
the normal state, i.e. $V_{dc}(B)=R_{N}I_{dc}$. This means that
the transition of the structure as a whole from the state close to
superconducting to the state close to normal and back can occur at
certain field and current values. The $V_{dc}(B)$ oscillations
more resemble abrupt jumps between the superconducting and normal
states as magnetic field changes.

We believe that the great amplitude of the $V_{dc}(B)$
oscillations can be due to the periodical magnetic-field-dependent
overheating of the nonequlibrium region by TAPSC up to the
effective quasiparticle temperature slightly above $T_{c}$ and
then by a following cooling of the overheated region to the
temperature slightly below $T_{c}$.

\section{CONCLUSION}
We found an unexpected interaction (nonlinear coupling) between
quantum magnetic-resistive states of two superconducting loops in
series with a very large distance between the loop centers close
to the penetration depth of a nonuniform electric field into a
superconductor $\Lambda_{E}>>\xi (T)$. Note that according to the
present-day studies such an interaction should not be expected in
the superconducting aluminum structure of two asymmetric circular
loops (Fig. \ref{image}).

To detect the interaction we measured both quantum oscillations of
a rectified time-averaged direct voltage $V_{rec}(B)$ and a dc
voltage $V_{dc}(B)$ in the 2-loop structure. The $V_{rec}(B)$ and
$V_{dc}(B)$ curves were recorded for the structure biased only
with an alternating current (without a dc component) and only with
a direct current (without an ac component), respectively, with the
maximum current values close to critical and $T$ slightly below
$T_{c}$.

Detailed analysis of the oscillations shows that Fourier spectra
of $V_{rec}(B)$ and $V_{dc}(B)$ functions contain the sum and
difference of the loop fundamental frequencies, which implies an
interaction (nonlinear coupling) between the quantum states of two
loops. We believe the coupling most likely realizes through a
common bias ac (dc) due to nonequilibrium nonlocal effects. The
large value of the nonequilibrium length $\Lambda_{E}$ in the
structure allow us to observe the coupling between the quantum
states of successive loops at a considerable distance between
them.

Quasiparticle overheating of the structure should be expected to
increase in the effective nonequilibrium length and the loop
coupling. Moreover, the quasiparticle overheating would result in
a great increase in the oscillation amplitude.

Earlier in Refs. \cite{sit1, sit7, sit8}, it was suggested that
measurements of a rectified voltage $V_{rec}(B)$ can be used to
determine quantum magneto-resistive states of an asymmetric
circular loop and two directly connected asymmetric circular loops
\cite{sit1, sit7, sit8}. In this work, we found that measurements
of a dc voltage $V_{dc}(B)$ can be also used to describe the
quantum states of the asymmetric structure.

The spectra of $V_{rec}(B)$ and $V_{dc}(B)$ oscillations are
complicated. Apart from the fundamental frequencies of both loops,
summation and difference fundamental frequencies, the spectra
contain low and extra combination frequencies. The extra
combination frequencies are linear combinations of the loop
fundamental frequencies and low frequencies.

The $V_{rec}(B)$ and $V_{dc}(B)$ oscillations in the structure
radically differ from the LP oscillations.

Quasiparticle overheating of the structure should be taken into
consideration when the $V_{rec}(B)$ and $V_{dc}(B)$ oscillations
are analyzed.

Further investigations would help to elucidate the nature of the
$V_{rec}(B)$ and $V_{dc}(B)$ oscillations.

\section{ACKNOWLEDGMENTS}
We thank Yu. Khanin and P.~Shabelnikova for technical help. This
work was financially supported in the frame of the program of
fundamental investigations of DNIT RAS "Organization of
computations on new physical principles" and the program of RAS
Presidium "Quantum Macrophysics" (section "Mesoscopics").

\end{document}